\documentclass[11pt]{article}
\usepackage{amsmath,amsthm,comment}
\usepackage{algorithm}
\usepackage{algorithmic}
\usepackage{amsfonts}
\usepackage{graphicx}
\usepackage{multirow}
\usepackage{booktabs}
\usepackage[top=1in, bottom=1in, left=1in, right=1in]{geometry}
\usepackage{subfigure}
\usepackage{enumerate}
\usepackage{dsfont}
\usepackage{bm}
\usepackage{color}
\usepackage{caption}
\usepackage[colorlinks=true,citecolor=blue,linkcolor=blue,urlcolor=blue]{hyperref}
\allowdisplaybreaks
\numberwithin{equation}{section}
\theoremstyle{definition}

\theoremstyle{remark}

\newcommand{\bx}{\mathbf{x}}
\newcommand{\bX}{\mathbf{X}}

\newcommand{\balpha}{\boldsymbol{\alpha}}

\begin{document}
\title{EVIboost for the Estimation of Extreme Value Index under Heterogeneous Extremes}
\author{
	Jiaxi Wang\footnote{\scriptsize School of Data Science, Fudan University.
	Email: {jxwang019@gmail.com}}\quad
	Yanxi Hou\footnote{\scriptsize School of Data Science, Fudan University.
	Corresponding author, Email: {yxhou@fudan.edu.cn}}\quad
	Xingchi Li\footnote{\scriptsize Department of 
	Statistics, Texas 
	A\&M University.
	%College Station, TX 77843, U.S. 
	Email: {anthony.li@stat.tamu.edu}}\quad
	Tiandong Wang\footnote{\scriptsize Department of 
	Statistics, Texas 
	A\&M University.
	%College Station, TX 77843, U.S. 
	Email: {twang@stat.tamu.edu}}
}
\date{}
\maketitle
\begin{abstract}

Modeling heterogeneity on heavy-tailed distributions under a regression framework is challenging, and classical statistical methodologies usually place conditions on the distribution models to facilitate the learning procedure. However, these conditions are likely to overlook the complex dependence structure between the heaviness of tails and the covariates. Moreover, data sparsity on tail regions also makes the inference method less stable, leading to largely biased estimates for extreme-related quantities. This paper proposes a gradient boosting algorithm to estimate a functional extreme value index with heterogeneous extremes. Our proposed algorithm is a data-driven procedure that captures complex and dynamic structures in tail distributions. We also conduct extensive simulation studies to show the prediction accuracy of the proposed algorithm. In addition, we apply our method to a real-world data set to illustrate the state-dependent and time-varying properties of heavy-tail phenomena in the financial industry.

\end{abstract}


\section{Introduction}

Estimation of the {\it extreme value index} is one important problem in extreme value theory. Suppose the univariate observations $\{y_i\}_{i=1}^n$ are generated from an unknown distribution $F$. We say that $F$ lies in the maximum domain of attraction $\mathcal D_\gamma$ with an extreme value index $\gamma\in\mathbb R$ (write $F\in\mathbb D_\gamma$), if there exist sequences $a_n$ and $b_n>0$, and a nondegenerate distribution $G_\gamma$ such that 
\[
	\lim_{n\to\infty}\mathbb P\left(\frac{\max_{1\le i\le n}y_i-a_n}{b_n}\le y\right)= G_\gamma(y),\quad 1+\gamma y>0,
\]
for every continuity point $y$ of $G_\gamma$, and $G_\gamma(y)=\exp\left(-(1+\gamma y)^{-1/\gamma}\right)$. Here the extreme value index, $\gamma$, represents the heaviness of the distributional tail, $1-F$, and is a crucial feature when learning the heavy-tail phenomena.

For example, in \cite{dekkers1989}, high quantiles of distribution are estimated based on the estimators of the extreme value index and intermediate quantiles, where the extrapolation technique is applied. In \cite{gencay2003}, the performance of the generalized Pareto distribution (GPD) model is compared with other well-known methods for the estimation of value-at-risk at high-risk levels, showing that the GPD model is more robust for the estimation of high quantiles. In their studies, the estimation of the {\em tail index}, which is defined as the reciprocal of extreme value index, is the key to fitting a GPD model.

Classical methods for univariate models are established on the assumption that univariate observations are independent and identically distributed (i.i.d). This assumption leads to a common, homogeneous extreme value index, $\gamma$, of the tail distributions for extreme values.
%, which paves a convenient way to establish estimators and their asymptotic properties. 
The value of $\gamma$ can be classified into three types of tail behaviors. In general, a positive $\gamma$ indicates a heavy right tail of the underlying distribution while a negative $\gamma$ indicates the existence of an endpoint on the right tail; when $\gamma$ equals zero, the tail behavior is even more complicated. Therefore, it is necessary to take the range of the extreme value index into account before the estimation. In the literature, one famous estimator is the Hill estimator \cite{hill1975} for $\gamma > 0$. Other estimators include the Pickands estimator for general $\gamma\in\mathbb R$, the maximum likelihood estimator for $\gamma>-1/2$; see Chapter 3 of \cite{haan2006} for detailed discussion. 

Because of the constant extreme value index, it is difficult to extend the estimation problem to non-i.i.d cases. Researchers have made a great effort to generalize the i.i.d. assumption of the underlying distributions in two different directions. One is the so-called {\it heteroscedastic extremes}, where the underlying distributions for the observations are independent but not identical. However, a constant extreme value index is usually assumed for a benchmark distribution, and a tail equivalence condition is used between the distributions of observations and the benchmark. The tail equivalence condition implies that all the distributions share a common tail heaviness as the benchmark distribution. Under heteroscedastic extremes, \cite{einmahl2016} develops a variant for the classical Hill estimator and shows that its asymptotic properties are similar to the traditional Hill estimator. In addition, \cite{xu2020} further develops the regression framework for heteroscedastic extremes and applies it to the prediction of conditional expectiles with extreme levels. 

The other direction is called {\it heterogeneous extremes}, where the underlying distributions are still independent and not identical, but the tail equivalence condition is released. Instead, each distribution of observations may have its extreme value index, which implies the tail heaviness is heterogeneous across all observations. Since the number of extreme value indices equals the sample size, the estimation problems become much more complicated than heteroscedastic extremes. To the best of our knowledge, there is no theoretical result so far for the model based on heterogeneous extremes. However, under a regression framework of $\{(y_i,\bx_i)\}_{i=1}^n$, it is proper to assume the extreme value index $\gamma(\bx)$ as a function of the predictor $\bx$. Based on this extension, it is possible to assume some explicit or implicit forms of $\gamma(\bx)$, and parametric or nonparametric methods can be applied under the regression framework. For example, \cite{wang2009} assumes an exponentially linear form of a positive tail index regression (TIR) model and applies the maximum likelihood method to estimate the coefficients. In this paper, we focus on modeling positive $\gamma(\bx)$. Thus, the TIR model represents a typical parametric model for the extreme value index, which serves as a benchmark for comparison purposes in our study. 
%In summary, under the regression framework, the main task is to predict $\gamma(\bx)$ precisely given any $\bx$. 

This paper aims to provide a gradient tree-boosted algorithm for the nonparametric estimation of $\gamma(\bx)$ based on heterogeneous extremes, which we refer to as EVIboost in the rest of the paper. Gradient boosting is one of the most influential machine learning algorithms; see the influential works \cite{friedman2001} and \cite{zhang2005}. Our proposed EVIboost algorithm is motivated by the nonparametric regression and classification problems in both fields of statistics and machine learning. Its advantages are two-fold. On one hand, EVIboost is a data-driven approach, and there is no assumption on the parametric form of $\gamma(\bx)$, which is more realistic for many practical problems. In our simulation studies, we focus on the comparison between the EVIboost and TIR model in \cite{wang2009} to show the superior performance of the proposed method over statistical models.
On the other hand, the EVIboost is a robust learning algorithm for the tail region of data distributions. It is well known that the prediction of $\gamma(\bx)$ is not reliable due to the data sparsity on the tail region, and many statistical approaches usually focus on the center region of the data based on central limit theory. The EVIboost receives little influence of the bias on the tail region than statistical methods, making it practical to estimate a functional extreme value index. 

We organize our paper as follows. Section~\ref{sec:method} presents the proposed EVIboost algorithm, and in particular, Section~\ref{subsec:algo} combines the regular variation condition in extreme value theory with the gradient boosting algorithm, which produces the fundamentals of our EVIboost algorithm. The rest of Section~\ref{sec:method} further describes several issues on the selection of threshold, choice of tuning parameters as well as model interpretation. In Section~\ref{sec:sim}, we conduct a series of simulation studies and compare the performance of EVIboost with the maximum likelihood method as employed in the TIR model. Section~\ref{sec:data} is a real-world data analysis that applies our EVIboost algorithm to heavy-tailed financial data.

\section{Methodology}\label{sec:method}

\subsection{EVIboost for Heterogeneous Extreme Value Index}\label{subsec:algo}

Suppose the observations $\{(y_i,\bx_i)\}_{i=1}^n$ are independent copies of $(Y,\bX)\in\mathbb R\times\mathbb R^p$. Let $F_Y(y|\bx)=\mathbb P(Y\le y|\bX=\bx)$ be the conditional distribution of $Y$ given $\bX=\bx$. We further assume that the conditional distribution are in the maximum domain of attraction $\mathcal D_{\gamma(\bx)}$ with $\gamma(\bx)>0$ for all $\bx$, where $\gamma(\bx)$ is the extreme value index depending on the predictor $\bx$. Thus, $F_Y(\cdot|\bx)\in\mathcal D_{\gamma(\bx)}$ implies that there exists a slowly varying function $L(y|\bx)$ satisfying $L(y|\bx)\to\infty$ and $L(yt|\bx)/L(y|\bx)\to 1$ as $t\to\infty$ for any $y>0$ and $\bx$, such that as $y\to\infty$,
\begin{equation}\label{eq:taildist}
1-F_Y(y|\bx)= y^{-1/\gamma(\bx)}L(y|\bx)(1+o(1)).
\end{equation}
Here we follow the approach of \cite{hall1982} and assume that
\begin{equation}\label{eq:slowvary}
L(y|\bx)=c_0(\bx)+c_1(\bx)y^{-\beta(\bx)}+O(y^{-\beta(\bx)}),
\end{equation}
where $c_0(\bx)$, $c_1(\bx)$ and $\beta(\bx)$ are functions of $\bx$ with $c_0(\bx)$, $\beta(\bx)>0$. Furthermore, denote $L_1(y|\bx)=\partial L(y|\bx)/\partial y$ which converges to $0$ as $y\to\infty$. The tail conditional probability function can be derived as
\begin{equation}\label{eq:tailden}
f_Y(y|\bx) =\left(\frac{1}{\gamma(\bx)}y^{-(1/\gamma(\bx)+1)}L(y|\bx)-y^{-1/\gamma(\bx)}L_1(y|\bx)\right)(1+o(1)).
\end{equation}
Based on the properties of $L(\cdot|\bx)$ and $L_1(\cdot|\bx)$, we have that for any given $\bx$,
\[
	f_Y(y|\bx)\gamma(\bx)y^{(1/\gamma(\bx)+1)}\to c_0(\bx),\quad y\to\infty,
\]
which implies that the tail conditional probability density function can be well approximated by $\frac{c_0(\bx)}{\gamma(\bx)}y^{-(1/\gamma(\bx)+1)}$. However, to implement statistical approaches, it is necessary to applying a sufficiently large threshold $u_n$ to select those data with $y_i>u_n$. Thus, we can define a loss function by considering a transformed negative likelihood function and omitting some constants which are not related to $\gamma(\bx)$,
\begin{equation}\label{eq:loss}
\mathcal L_n(\gamma(\cdot)|u_n)=\sum_{i=1}^n\Psi(y_i,\gamma(\bx_i)|u_n)=\sum_{i=1}^n\left(\frac{\log(y_i/u_n)}{\gamma(\bx_i)}+\log(\gamma(\bx_i))\right)I(y_i>u_n),
\end{equation}
and given $u_n$, we intend to estimate
\begin{equation}\label{eq:opt}
\gamma^*_n(\cdot):=\gamma^*_n(\cdot|u_n) = \text{argmin}_{\gamma(\cdot)\in\mathcal F}\sum_{i=1}^n\Psi(y_i,\gamma(\bx_i)|u_n),
\end{equation}
where $\mathcal F$ is a class of functions. 

For the gradient boosting, a greedy stagewise algorithm described in \cite{friedman2001} assumes each candidate $F$ to be an ensemble of $M$ base learners,
\[
\gamma^{[m]}(\bx) = \gamma^{[0]}+ \sum_{i=1}^M\beta_mh(\bx|\balpha_m),
\]
where $h(\cdot|\balpha_m)$ is a base learner belonging to a class of simple functions with parameters $\balpha_m$, and $\gamma^{[0]}$ and $\beta_m$ are constants.

The initial estimate of $\gamma^*_n(\cdot)$ is the Hill estimator without including $\bx_i$, which can be obtained by using a constant $\theta$ in $\Psi$:
\begin{equation}\label{eq:initial}
\gamma^{[0]}=\text{argmin}_{\theta}\sum_{i=1}^n\Psi(y_i,\theta|u_n)=\frac{1}{k_n}\sum_{i=1}^n\log\left(y_i/u_n\right)I(y_i>u_n),
\end{equation}
where $k_n=\sum_{i=1}^nI(y_i>u_n)$ is the number of $y_i$ above the threshold $u_n$. 

Then, by the gradient boosting algorithm,  at the $m$-th step, the current update is
\begin{equation}\label{eq:update}
\gamma^{[m]}(\bx) = \gamma^{[m-1]}(\bx) + \beta_mh(\bx|\balpha_m),\quad m=1,2,\ldots,M.
\end{equation}
The current negative gradient is
\begin{equation}\label{eq:neggrad}
\tilde y_i^{[m]} = -\frac{\partial\Psi(y_i,\gamma(\bx_i)|u_n)}{\partial\gamma(\bx_i)}\Big|_{\gamma(\bx_i)=\gamma^{[m-1]}(\bx_i)} = \left(\frac{\log(y_i/u_n)-\gamma^{[m-1]}(\bx_i)}{\gamma^{[m-1]}(\bx_i)^2}\right)I(y_i>u_n).
\end{equation}
We use an L-terminal node regression tree as the base learner such that
\[
	h(\bx|\{b_l,R_l\})=\sum_{l=1}^Lb_lI(\bx\in R_l).
\]
Then, we apply the algorithm in \cite{friedman2001} to split the regions $R_{lm}$, and \eqref{eq:update} becomes 
\begin{equation}\label{eq:update2}
\gamma^{[m]}(\bx) = \gamma^{[m-1]}(\bx) +\sum_{l=1}^L\eta_{lm}I(\bx\in R_{lm}),
\end{equation}
with the coefficient $\eta_{lm}= \beta_m\text{ave}_{i:\bx_i\in R_{lm}}\tilde y_i^{[m]}$. Thus, the optimal coefficients are the solution to
\begin{equation}\label{eq:basecoef}
\eta_{lm} =\underset{\eta}{\text{argmin}}\sum_{i:\bx_i\in R_{lm}}\Psi(y_i,\gamma^{[m-1]}(\bx_i)+\eta),\quad l=1,2,\ldots,L,
\end{equation}
which has no explicit solution. Therefore, a single Newton-Raphson step is applied, as described in \cite{friedman2001}. This leads to the following result
\begin{equation}\label{eq:basecoef2}
\eta_{lm} =\frac{\sum_{i:\bx_i\in R_{lm}}\tilde y_i^{[m]}}{\sum_{i:\bx_i\in R_{lm}}\left(2\tilde y_i^{[m]}/\gamma^{[m]}(\bx_i)+1/\gamma^{[m]}(\bx_i)^2I(y_i>u_n)\right)}
\end{equation}
where $k_{n,lm}=\sum_{i:\bx_i\in R_{lm}}$. By adding a shrinkage factor $\nu\in(0,1]$ to \eqref{eq:update2} and update the current estimate at each region $R_{lm}$, we have that
\begin{equation}\label{eq:update3}
\gamma^{[m]}(\bx) = \gamma^{[m-1]}(\bx) +\nu\eta_{lm}I(\bx\in R_{lm}),\quad l=1,2,\ldots, L.
\end{equation}
where $\nu$ is the shrinkage factor which controls the rate of learning. Thus, after $M$ iterations, it turns out the estimator $\gamma^{[M]}(\cdot)$ of $\gamma^*_n(\cdot)$ for \eqref{eq:opt}. The EVIboost algorithm is summarized in Algorithm \ref{alg:tb}.

\begin{algorithm}[htbp]
\caption{The EVIboost algorithm for estimation of $\gamma(\cdot)$.}\label{alg:tb}
\begin{algorithmic}[1]
\STATE {\bf Initialize} $\gamma^{[0]}$
	\[
		\gamma^{[0]}=\frac{1}{k_n}\sum_{i=1}^n\log\left(y_i/u_n\right)I(y_i>u_n).
	\]
\STATE {\bf For} $m=1,2,\ldots,M$, {\bf do} 
\begin{itemize}
\item[a.] Compute the negative gradient $(\tilde y_1^{[m]},\tilde y_2^{[m]},\ldots,\tilde y_n^{[m]})^T$,
\[
\tilde y_i^{[m]} = \left(\frac{\log(y_i/u_n)-\gamma^{[m-1]}(\bx_i)}{\gamma^{[m-1]}(\bx_i)^2}\right)I(y_i>u_n),\quad i=1,2,\ldots,n.
\]
\item[b.] Fit $\{(\tilde y_i^{[m]},\bx_i)\}_{i=1}^n$ to an $L$-terminal node regression tree,
\[
\{R_{lm}\}_1^L= L\text{-terminal node tree of }\{(\tilde y_i^{[m]},\bx_i)\}_1^n.
\]
\item[c.] Approximate the optimal terminal node predictions $\eta_{lm}$ of $R_{lm}$,
\[
\eta_{lm} =\frac{\sum_{i:\bx_i\in R_{lm}}\tilde y_i^{[m]}}{\sum_{i:\bx_i\in R_{lm}}\left(2\tilde y_i^{[m]}/\gamma^{[m]}(\bx_i)+1/\gamma^{[m]}(\bx_i)^2I(y_i>u_n)\right)},\quad l=1,2,\ldots,L.
\]
\item[d.] Update $\gamma^{[m]}(\bx)$ for each $R_{lm},\, l=1,2,\ldots,L$,
\[
\gamma^{[m]}(\bx) = \gamma^{[m-1]}(\bx) +\nu\eta_{lm}I(\bx\in R_{lm}),\quad l=1,2,\ldots, L.\]
\end{itemize}
\STATE {\bf End for} Return $\gamma^{[M]}(\bx)$ as the final estimate.
\end{algorithmic}
\end{algorithm}

\subsection{Choice of Tuning Parameters}\label{subsec:tuning}

To implement Algorithm~\ref{alg:tb}, we need to choose the threshold $u_n$ as well as three critical tuning parameters in advance: (1) the learning rate, $\nu$, (2) the number of trees, $M$, and (3) the number of terminal nodes, $L$. Here $\nu$ and $M$ together control the length and the total number of steps in the gradient boost optimization process, and $L$ specifies the complexity of an individual regression tree. For instance, $L=2$ indicates only one splitting variable in the tree; thus, the tree models the main effects of predictors.

For a given threshold $u_n$, we first select several discrete values for $\nu$ and $L$, then with values of $\nu$ and $L$ chosen, we apply a cross-validation (CV) method to tune the number of trees, $M$, aiming to minimize the loss function $\mathcal L_n(\gamma(\cdot)|u_n)$ on the validation set. Specifically, we adopt a $K$-fold CV approach and denote $\gamma^{[M;\nu,L]}_{(-j)}(\bx)$ as the estimator of $\gamma(\bx)$ with the $j$-th fold as the validation dataset and the other $K-1$ folds as training dataset. This gives the validation loss as
\[
	CV(M,\nu,L):= \mathcal L_n\left(\gamma^{[M;\nu,L]}_{(-j)}(\bx)|u_n\right).
\]
Given $(\nu, L)$, we select the optimal $M$ by minimizing the validation loss, i.e. 
\[
	\widehat M_{\nu,L}:=\text{argmin}_M  CV(M,\nu,L).
\]
Hence, the optimal choice of $(M,\nu,L)$ becomes,
\[
(\widehat\nu,\widehat L) := \text{argmin}_{\nu, L}CV(\widehat{M}_{\nu, L}, \nu, L),
\qquad\text{and}\qquad \widehat{M} := M_{\widehat\nu,\widehat L}.
\]

\subsection{Selection of the Threshold}\label{subsec:thres}

Another important parameter to determine is the threshold, $u_n$, which 
may largely affect the asymptotic distribution \eqref{eq:taildist}.
%Here we distinguish the choice of the threshold, $u_n$, from that of the tuning parameters since the selection of threshold  
Here we choose the threshold $u_n$ so that it controls the essential sample size on tail region. Since \eqref{eq:taildist} implies that for any given $\bx$,
\begin{equation}\label{eq:tailden2}
\lim_{t\to\infty}\frac{1-F_Y(ty|\bx)}{1-F_Y(t|\bx)}=y^{-1/\gamma(\bx)},\quad y>0,
\end{equation}
we define $\tilde U_i:=(y_i/u_n)^{-1/\gamma^{[\widehat M;\hat\nu,\widehat L]}(\bx)}$ for given values of $\bx$, and let $F_n$ be the empirical distribution of $\tilde U_i$ for which $y_i>u_n$. The optimal choice of $(\widehat M,\hat\nu,\widehat L)$ is determined under the given $u_n$. 
Following methods in \cite{clauset:2009} and \cite{wang2009}, we consider three different discrepancy measures:
%In addition to the discrepancy measure $D_1$ in , we also consider Kolmogorov-Smirnov $D_2$ and Anderson-Darling $D_3$ discrepancy measures as follows.
\begin{align}
D_1(u_n)&:=\frac{1}{k_n}\sum_{y_i>u_n}(\tilde{U}_i-F_n(\tilde{U}_i))^2,\\
D_2(u_n)&:=\sup_{y_i>u_n}\lvert\tilde{U}_i-F_n(\tilde{U}_i)\rvert,\\
D_3(u_n)&:=\frac{1}{k_n}\sum_{y_i>u_n}\frac{(\tilde{U}_i-F_n(\tilde{U}_i))^2}{\tilde{U}_i(1-\tilde{U}_i)},
\end{align}
where $D_1(\cdot)$ is identical to the setting in \cite{wang2009}, and $D_2(\cdot)$, $D_3(\cdot)$ correspond to the Kolmogorov-Smirnov and Anderson-Darling distances, respectively. Then $u_n$ can be determined by minimizing a selected discrepancy measure.

%{\color{blue}{Notation is weird. Why we need $q$?}}{\color{red}Give a clear definition between $q$ and $u_n$} {\color{cyan} The reason why we introduce $q$ is that, it's easier to select the fractile that $\mu_n$ corresponds to, than to select $\mu_n$ directly. I rewrite the paragraph below to make it clearer.}

A feasible approach to select $u_n$ is by determining the tail sample fraction $q=k_n/n$, which we can consider as a turning parameter, describing the proportion of $y_i$ exceeding $u_n$ used in the algorithm. Then, the selection of the threshold is equivalent to the determination of the tail fraction. Given the response $y_1,\ldots,y_n$, we consider a finite sequence of tail fractions $\{q_s\}_{s=1}^S$ equally spaced on the interval $[0,1]$ (e.g. $\{0.01,0.02,\ldots,0.99\}$). For each $q_s$, let $u_{sn}$ be the corresponding $(1-q_s)$-th sample quantile of $y_1,\ldots,y_n$. Then the optimal threshold is 
\begin{equation}\label{eq:optthres}
u_n^*=\underset{u_n\in\{u_{sn}\}_{s=1}^{S}}{\text{argmin}}\, D(u_{sn}).
\end{equation}
and the final estimate of $\gamma(\cdot)$ is $\gamma^{[\widehat M;\hat\nu,\widehat L]}(\cdot)$ given $u_n^*$.

\subsection{Model Interpretation}
%{\color{blue} We need to discuss the writing of this section.}{\color{red} Write it in a strict way. Refer to other papers how to write the model interpretation.} {\color{cyan} I mainly refer to three papers, including Grabit (Fabio, 2019), TDboost (Yang, 2018) and gradient boosting for quantile regression (Jasper, 2021), and make some modifications on the last version.}

%{\color{cyan}{Compared to other nonparametric statistical learning algorithms, the EVIboost uses regression trees as base learners, which enables us to provide more interpretations.}} {\color{blue}{any refs?}} {\color{blue} This statement is vague. Not sure what it means.}  {\color{cyan} Actually this statement is referred to TDboost.}

The tree-based models are more interpretive than other nonparametric machine learning algorithms such as neural networks and support vector machines. In this subsection, we briefly discuss two tools, the (feature) importance measure and the partial dependence plot.

In many real applications, one usually wants to identify the importance of features/covariates on the predictions of the interesting objects. Here we apply an impurity-based method proposed by \cite{breiman1984} to evaluate the importance of each individual feature $x_1,\ldots,x_p$. The definition starts within a single tree $T_{m}$. Suppose $x_i$ is the feature of interest, then the importance of $x_i$ on $T_m$ is given by 
$$ 
I_{m}(x_{i}) = \sum_{j=1}^{J} \Delta \delta_j\cdot \mathbb{I}(x_{i},j),\quad i=1,2,\ldots,p,
$$
where the sum is over all $J$ non-terminal nodes of $T_m$, and $\Delta \delta_j$ is the reduction of the squared error caused by node $j$ (recall that during the construction of a tree, the algorithm greedily searches for a split that can maximize the reduction in MSE). The indicator $\mathbb{I}(x_{i},j)$ equals one if the node $j$ uses $x_{i}$ to split and zero otherwise. Since the EVIboost is an ensemble of $M$ trees, we take the average of $I_{1}(x_{i}),\ldots, I_{M}(x_{i})$ as the importance measure of $x_i$, i.e.
$$ I(x_{i})=\frac{1}{M}\sum_{m=1}^{M}I_{m}(x_{i}).$$
Different from the exponentially linear form as in the TIR model \cite{wang2009}, the importance measure $I(x_{i})$ considers not only the main effects but also interactions among variables. However, $I(x_{i})$ may be biased so that a feature irrelevant to the response may still have non-zero importance if it is chosen as the splitting variable by any nodes; see \cite{white1994} and \cite{zuc2008}. Therefore, we follow the methods in \cite{yang2018} to derive a modified importance measure. Let $\textbf{x}$ be the $n\times p$ design matrix, repeat steps (1) to (3) for $r$ from 1 to $R$.
\begin{enumerate}[(1)]
\item Generate an $n\times p$ matrix $\textbf{z}^{(r)}$ by randomly shuffling the $n$ rows of $\textbf{x}$, while the order of columns are unchanged. Bind $\textbf{x}$ and $\textbf{z}^{(r)}$ by columns, then denote the $n\times 2p$ matrix $[\textbf{x},\textbf{z}^{(r)}]$ as $\textbf{x}^{(r)}$.
\item Implement the EVIboost model using $\{y,\textbf{x}^{(r)}\}$.
\item Compute the importance measures $I^{(r)}(x_{i})$ for $x_{i}$ and $I^{(r)}(z^{(r)}_{i})$ for $z^{(r)}_{i}$ respectively, where $z^{(r)}_{i}$ is the $i$th column of $\textbf{z}^{(r)}$. 
\end{enumerate}
Since the pseudo-predictor $z^{(r)}_{i}$ is shuffled from $x_i$, $z^{(r)}_{i}$ has the same number of possible splits as $x_i$, and is equivalently possible of being selected by tree nodes. Hence, we take $I^{(r)}(z^{(r)}_{i})$ as a bias approximation for $I(x_{i})$, and the modified importance measure for $x_{i}$ is given by
$$
I^{\star}(x_i)=\frac{1}{R}\left(\sum_{r=1}^{R}I^{(r)}(x_{i})-\sum_{r=1}^{R}I^{(r)}(z^{(r)}_{i})\right).
$$
%{\color{blue} Summarize the method. I presume this is the relative importance measure which is being referred to later??} {\color{red} write out what we used.} {\color{cyan}It is similar to relative importance measure since it uses random shuffle but not the same. I write the method in detail here.}

One limitation of the importance measure is that it cannot demonstrate how the estimated function $\gamma^{[M;\nu,L]}(\cdot)$ varies along with the features. To solve this problem, we employ the partial dependence plots introduced in \cite{friedman2001}. We divide the predictors $\bX$ and its observation $\bx_i$ into two non-overlapping subsets $\bX_s$ and $\bX_{-s}$, and $\bx_{i,s}$ and $\bx_{i,-s}$, where $s$ is a nonempty index subset of $\{1,2,\ldots,p\}$. The partial dependence of $\bX_s$ at $\bx_s$ is then given by
\begin{equation}
\label{eq:partial_dep}
\bar{\gamma}_s(\bx_s)=\frac{1}{n}\sum_{i=1}^{n}\gamma^{[M;\nu,L]}(\bX_s=\bx_s,\bX_{-s}=\bx_{i,-s}).
\end{equation}
Note that when $\bX_s$ are independent of $\bX_{-s}$, $\bar{\gamma}_s(\cdot)$ will serve as an estimator of the conditional expectation $E(\hat{\gamma}(\bx_s)|X_s=\bx_s)$. To fully depict the marginal effects of $\bX_s$ with respect to $\gamma^{[M]}(\cdot)$, we then plot $\bar{\gamma}_s(\cdot)$ versus domain of $\bX$. In the real data analysis, Figure~\ref{fig:pd} illustrates the applications of the modified importance measure and the one-dimensional partial dependence plot.

%{\color{blue} Then what?}{\color{red} Give intuition and explanation about partial dependence.}{\color{cyan} I made a clearer explanation.}

\section{Simulation Study}\label{sec:sim}

\subsection{Simulated Models}\label{subsec:sim}

In this section, we conduct simulation studies to compare the prediction accuracy of our EVIboost algorithm with the TIR model in \cite{wang2009}. We follow a similar simulation setup as in \cite{wang2009}, but we consider more models, including both exponentially linear and nonlinear forms. Note that TIR is restricted by an exponentially linear form of $\gamma(\cdot)$ whereas the proposed EVIboost algorithm possesses more model flexibility by allowing nonparametric functions.% Suppose $F_{Y}(y|\textbf{x})$ satisfies \eqref{eq:taildist}, and a fixed tail fraction $q$ is given thus the associated threshold can be determined.
We describe the scheme of generating simulated samples and calculating performance metrics as follows.

{\bf Step 0}: Determine parameters in the simulation. 

We choose $p=10$, $m=0.10$, $C=1/3$, $n=n^*=1000$ and $R=100$. Also, set the tail fraction $q=0.1,0.05,0.025$ and let $q^\star$ be the optimal fraction given by
\[
	q^\star :=\text{argmin}_q (D_1(u_n)+D_2(u_n)+D_3(u_n)),
\]
Note that $D_i(u_n)$, $i=1,2,3$, are the three discrepancy measures, and $u_n$ is the threshold given the tail fraction $q$. 

{\bf Step 1}: Generate a training sample $(\textbf{x}_{i}^{(r)},y_{i}^{(r)})$ of size $n$.
%such that $\bx_i^{(r)}$ is from $F_{X}$ and then $y_i^{(r)}$ is from  $F_{Y}(\cdot|\bx_i^{(r)})$. 

Simulate $\textbf{Z}_{i}:=(Z_{i,1},\ldots,Z_{i,p})\sim N_{p}$ with zero mean and $\text{Cov}(Z_{i,j}, Z_{i,k})= {\frac{1}{2}}^{|j-k|},\, j,k=1,\ldots,p$. Denote $x_{i,j}^{(r)}=2\sqrt{3}(\Phi(Z_{i,j})-\frac{1}{2})$, where $\Phi$ is the CDF of a standard normal random variable, and $x_{i,j}^{(r)}$ is the $j$-th coordinate of $\bx_i^{(r)}$.

Given $\bx_i^{(r)}$, we then simulate $y_{i}$ from
\[
F(y|\textbf{x}_{i}^{(r)})=1-\frac{(1+m)y}{y^{1/\gamma(\textbf{x}_{i}^{(r)})}+my},
\]
where $\gamma(\bx)$ is of one of the following forms:
\begin{enumerate}
    \item[(1)] $ \gamma_{1}(\textbf{x})=C\exp\left(-\frac{1}{2}x_{1}+\frac{1}{3}x_{2}-\frac{1}{3}x_{3}\right) $;
    \item[(2)] $ \gamma_{2}(\textbf{x})=C\exp\left(\frac{2}{p}\sum_{i=1}^{p}x_{i}\times(-1)^{i}\right)$;
    \item[(3)] $ \gamma_{3}(\textbf{x})=C\exp\left(-\frac{1}{2}x_{1}^{2}+\frac{1}{3}x_{2}^{2}-\frac{1}{3}x_{3}^{2} \right) $;
    \item[(4)] $ \gamma_{4}(\textbf{x})=\exp\left(-(x_{1}+x_{2})^{2}-(x_{2}+x_{3})^{4} \right) $;
    \item[(5)] $ \gamma_{5}(\textbf{x})=\exp\left(-\sqrt{x_{1}-x_{2}}-\frac{1}{\sqrt{x_{3}-x_{4}}} \right) $.
\end{enumerate}

{\bf Step 2}: Given the tail fraction $q$, set the value of $u_{n}^{(r)}$ to be the $(1-q)$-th sample quantile of $y_{1}^{(r)},y_{2}^{(r)},\ldots,y_{n}^{(r)}$.

{\bf Step 3}: Estimation of the extreme value index function $\gamma(\cdot)$.

Given the threshold $u_{n}^{(r)}$, we implement the EVIboost algorithm to obtain $\hat\gamma^{[M;\nu,L](r)}(\cdot)$ and apply the maximum likelihood estimation for TIR model in \cite{wang2009} to obtain $\hat\gamma_{TIR}(\cdot)$. Note that the tuning parameters $M,\nu,L$ of EVIboost are obtained by following a five-fold CV approach as described in Section~\ref{subsec:tuning}.

{\bf Step 4}: Calculate the mean squared error of predictions on a testing sample.

Generate a testing sample, $(\textbf{x}^{*(r)}_{i},y^{*(r)}_{i})$ of size $n^*$, in the same way as in Step 1. Then use the testing sample to evaluate the two estimators in Step 3 by a mean squared error:% of predictions such that 
\[
\hat\delta^{(r)}=\frac{1}{n^*}\sum_{i=1}^{n^*}\left(\hat\gamma(\textbf{x}^{*(r)}_{i})-\gamma(\textbf{x}^{*(r)}_{i})\right)^{2},
\]
where $\hat\gamma(\cdot)$ is either $\hat\gamma^{[M;\nu,L](r)}(\cdot)$ or $\hat\gamma_{TIR}(\cdot)$, and $\hat\delta^{(r)}_{EVI},\hat\delta^{(r)}_{TIR}$ denote $\hat\delta^{(r)}$ under the EVIboost and TIR models, respectively.

{\bf Step 5}: Repeat Steps 1 to 4 for $r$ from 1 to $R$ independently and compare the prediction performance of $\{\hat\delta^{(r)}_{EVI}\}_{r=1}^R$ and $\{\hat\delta^{(r)}_{TIR}\}_{r=1}^R$. 

We compare the prediction performance of the EVIboost and TIR models by sketching the boxplots of $\{\hat\delta^{(r)}_{EVI}\}_{r=1}^R$ and $\{\hat\delta^{(r)}_{TIR}\}_{r=1}^R$. To evaluate the advantage of our method quantitatively, we compute an efficiency ratio as
$ \frac{\text{med}(\{\hat\delta^{(r)}_{EVI}\}_{r=1}^R)}{ \text{med}(\{\hat\delta^{(r)}_{TIR}\}_{r=1}^R)}$, where $\text{med}(\cdot)$ means the sample median of a sequence.

%{\color{red} give a efficiency ratio formula.} {\color{cyan}{OK.}}

\begin{figure}[htbp]
\centering
\includegraphics[scale=0.56]{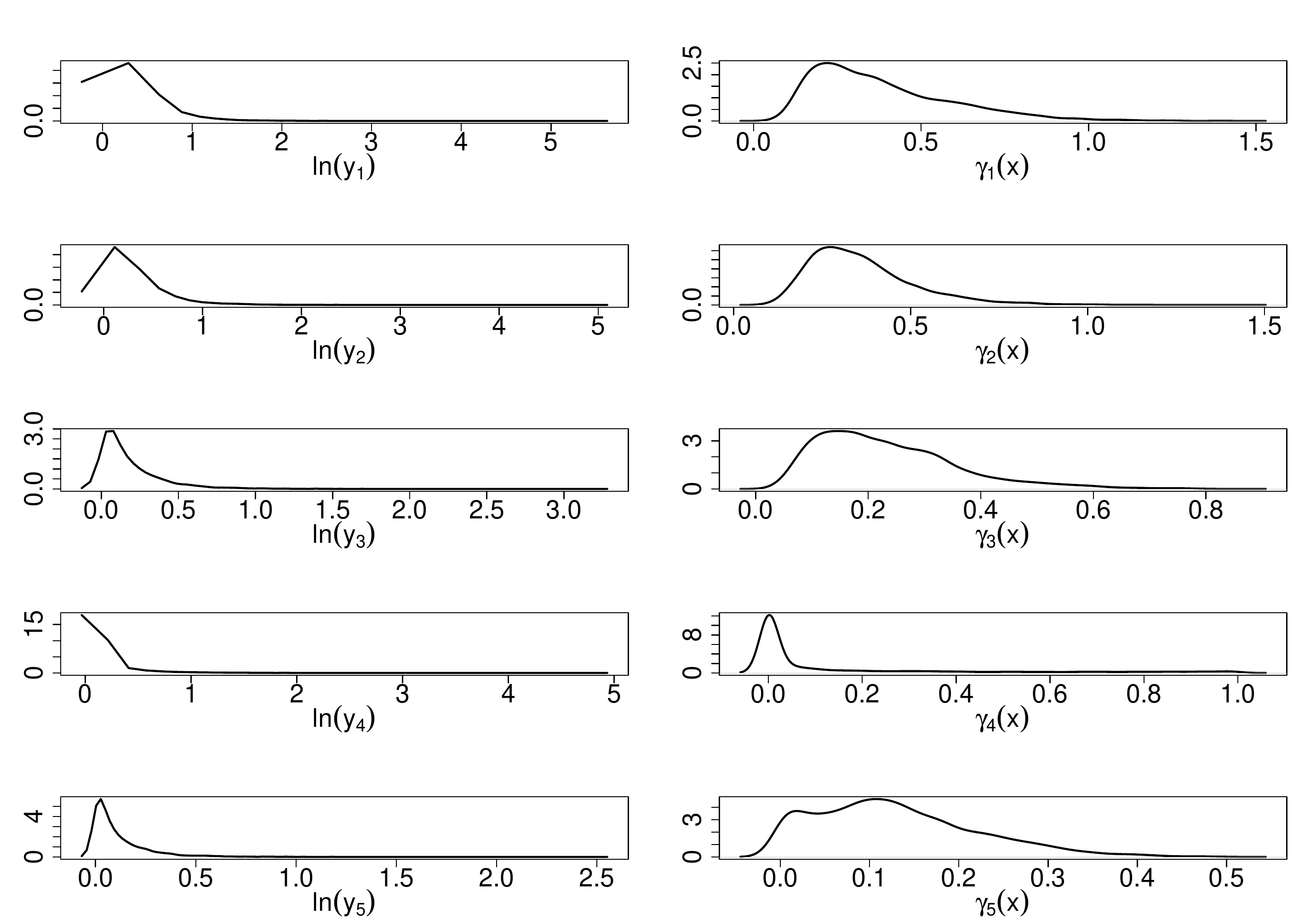}
\caption{Density plots of simulated $\log(y_{i})$ and$\gamma(\textbf{x}_{i})$.
%{\color{blue} In your graph, set $log='x'$, then you'll have the density plot for $y$, but your x-axis will be on log-scale. Plotting $log(y)$ directly looks weird.}
}
\label{fig:density}
\end{figure}

In Step 1, each marginal distribution of $\bx_i^{(r)}$ is $U[-\sqrt{3},\sqrt{3}]$ with unit variance, and any $x_i, x_j\ i\neq j$ is pairwisely correlated. %{\color{blue} pairwise correlated??} {\color{cyan}{Yes.}}. 
For the choice of $\gamma(\cdot)$, we assume $\gamma(\cdot)$ to be exponentially linear in Cases (1) and (2), which agrees with assumptions in the TIR model. However, in Case (1), only the first three covariates have impacts on $\gamma(\cdot)$, whereas in Case (2), all covariates are equally influential. Cases (3) to (5) assume exponentially nonlinear forms with higher-order interaction terms, under which the TIR model is misspecified. Figure~\ref{fig:density} gives the densities of simulated data in all cases. In Step 2, we intend to see the prediction performance of the EVIboost and TIR models given the upper tail fraction and the optimal one chosen by the TIR model.
%{\color{blue} Density plots for y are barely visible. Maybe switch to log-scale for the x-axis?} {\color{red} use $\log y_i $ on x-axis of the left plot in Figure 1.} {\color{cyan}{OK}}
%For generality, simulations are conducted under $q=0.1,0.05,0.025$ (we set $q$ to a low level thus $u_{n}$ can be sufficiently large to guarantee the approximation in \eqref{eq:tailden}) and an optimal fraction $q^{\star}$ selected by TIR. We range $q$ from $0.005$ to $0.995$ by $0.005$ and compute the three discrepancy measures $D_1 \sim D_3$ under each $q$. Then $q^{\star}$ is obtained by minimizing the sum of $D_1 \sim D_3$. We consider various forms of $\gamma(\textbf{x})$, as is listed below. 

\subsection{Results}

\begin{figure}[htbp]
  \centering
  \subfigure[]{\includegraphics[scale=0.41]{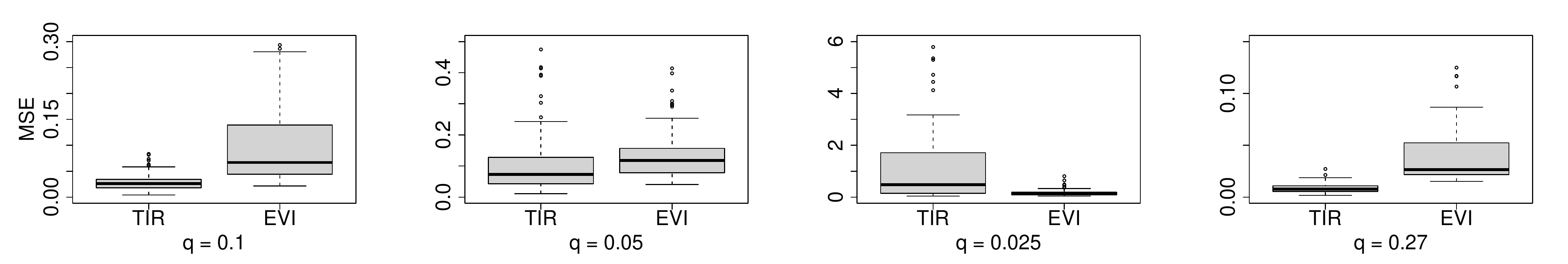}}\\
  \subfigure[]{\includegraphics[scale=0.41]{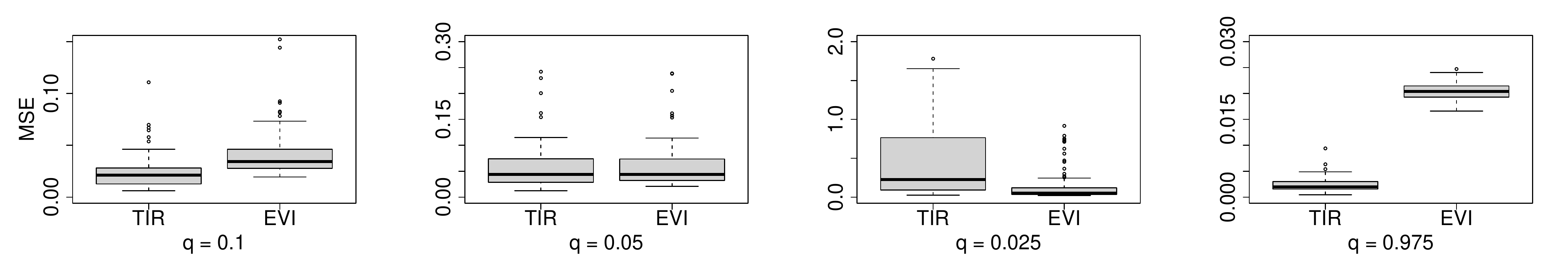}}\\
  \subfigure[]{\includegraphics[scale=0.41]{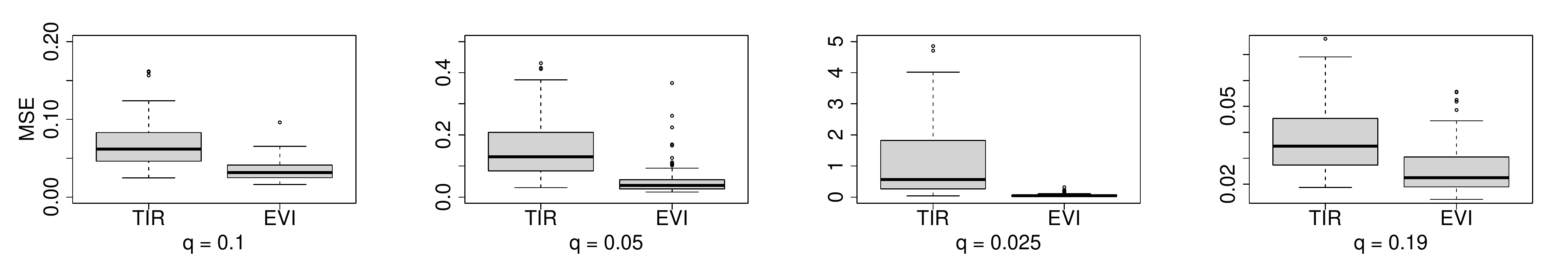}}\\
  \subfigure[]{\includegraphics[scale=0.41]{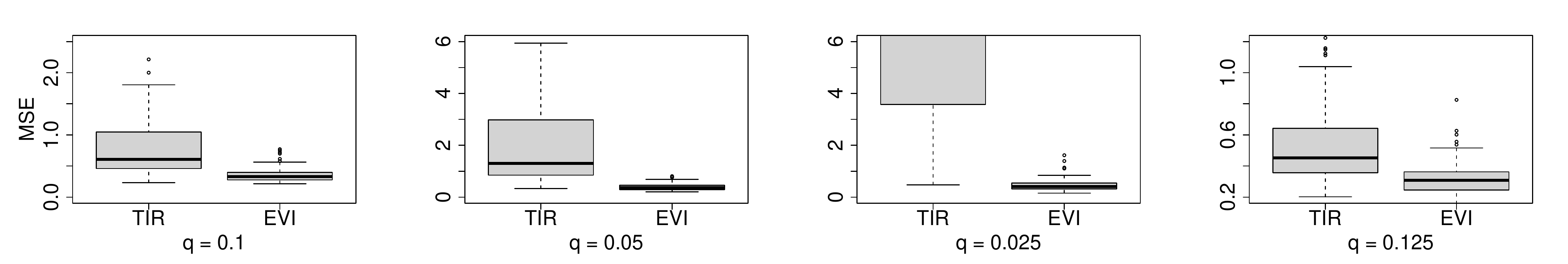}}\\
  \subfigure[]{\includegraphics[scale=0.41]{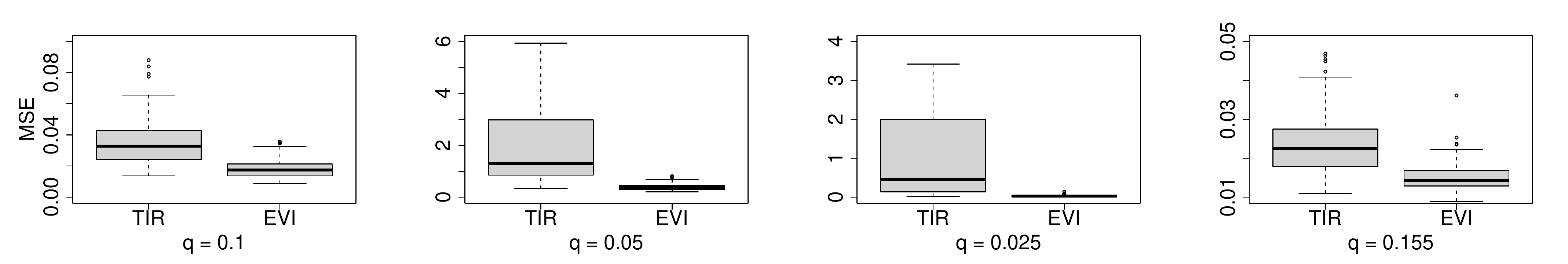}}\\
  \caption{Mean squared error (MSE) of TIR and EVIboost estimators under $R=100$ replications. Plots (a) to (e) correspond to Cases (1) to (5), respectively. %Some outliers are excluded in the boxplots for clearer comparison.
  }
\label{fig:simu}
\end{figure}

Now we discuss our numerical results from the simulation study.
Figure~\ref{fig:simu} presents the boxplots of the mean squared error of predictions in Section~\ref{subsec:sim} for both the EVIboost and TIR models across all five cases. 
When $\gamma(\textbf{x})$ is exponentially linear, i.e. in Cases (1) and (2), the prediction errors of EVIboost estimators are higher than those of TIR under low thresholds ($q=0.1$), but they are equivalent or even lower when $q=0.05, 0.025$ (see the middle two panels of Figure~\ref{fig:simu}(a)). As explained in Section \ref{subsec:sim}, Case (1) corresponds to the simulation model in \cite{wang2009}, where the TIR model correctly specifies $\gamma(\cdot)$. For the optimal fraction $q^\star$ chosen by the TIR model, the TIR estimators
outperform the EVIboost ones if $\gamma(\cdot)$ is correctly specified. However, the EVIboost performs much better than TIR when $\gamma(\textbf{x})$ is no longer exponentially linear, which shows that the proposed algorithm is a data-driven method. In Cases (3) to (5), the mean squared error of the EVIboost is consistently lower for all chosen fractions $q$. We also summarize the efficiency ratios at all fractions in Table~\ref{tab:summary}. 
%offers more details, in which the efficiency of EVI is evaluated by $$E= median: \{\delta_{EVI}^{(1)},\delta_{EVI}^{(2)}...\delta_{EVI}^{(R)}\} \div  median: \{\delta_{TIR}^{(1)},\delta_{TIR}^{(2)}...\delta_{TIR}^{(R)}\} $$
Overall, the accuracy of the EVIboost and TIR are both sensitive to the tail fraction $q$. However, EVIboost has a better performance when $q$ is small, where in contrast, the prediction results from the TIR model show significant biases and variations. 

\begin{figure}[htbp]
  \centering
  \subfigure[]{\includegraphics[scale=0.31]{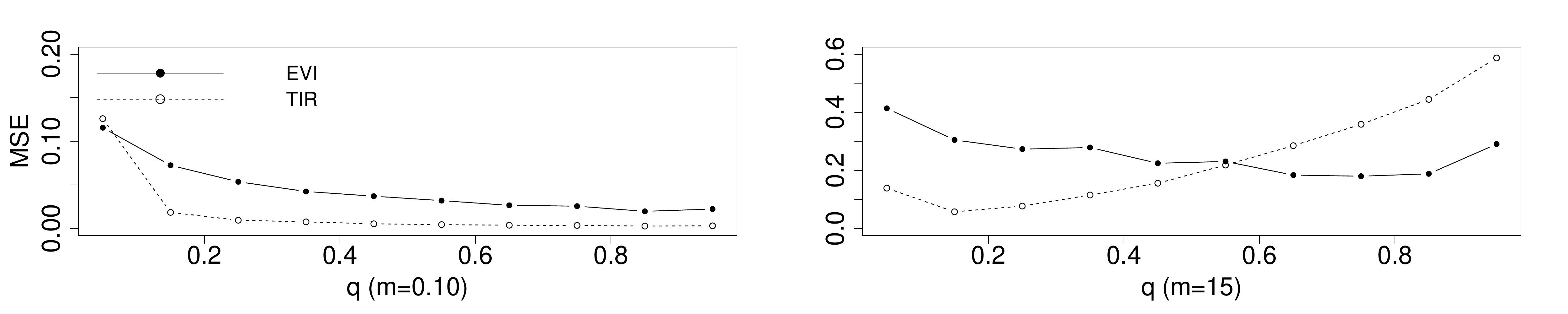}}\\
  \subfigure[]{\includegraphics[scale=0.31]{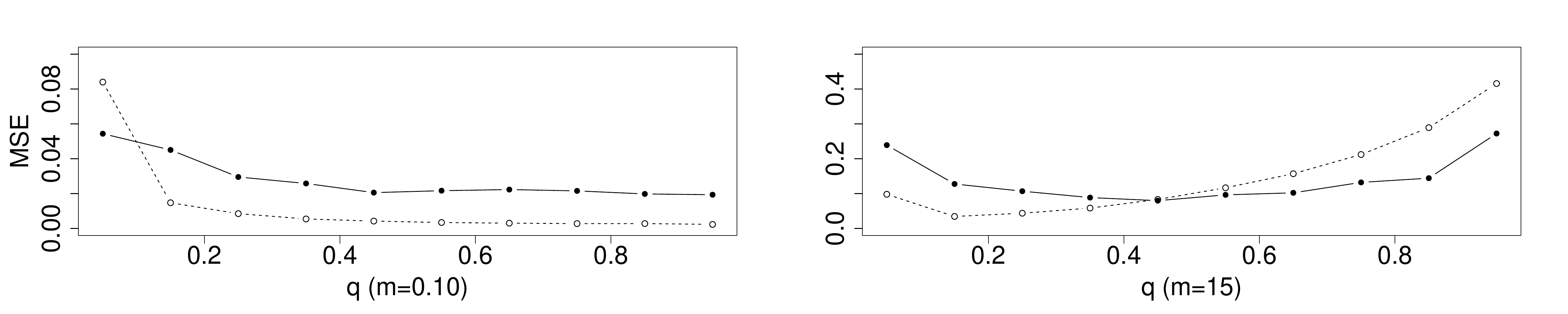}}\\
  \subfigure[]{\includegraphics[scale=0.31]{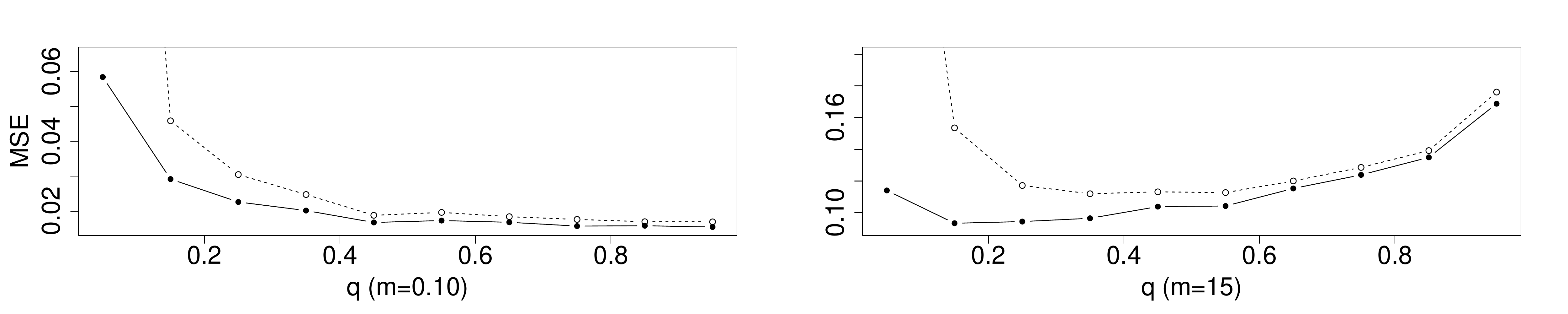}}\\
  \subfigure[]{\includegraphics[scale=0.31]{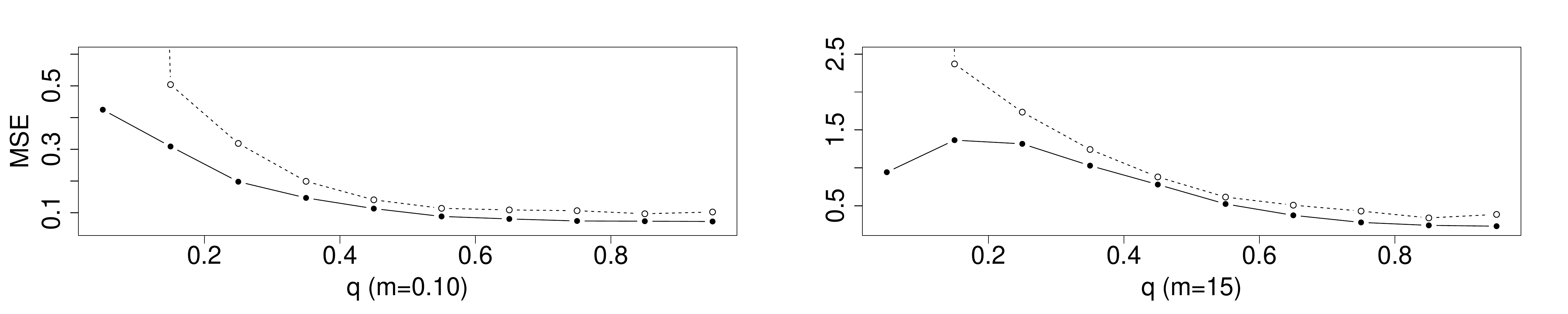}}\\
  \subfigure[]{\includegraphics[scale=0.31]{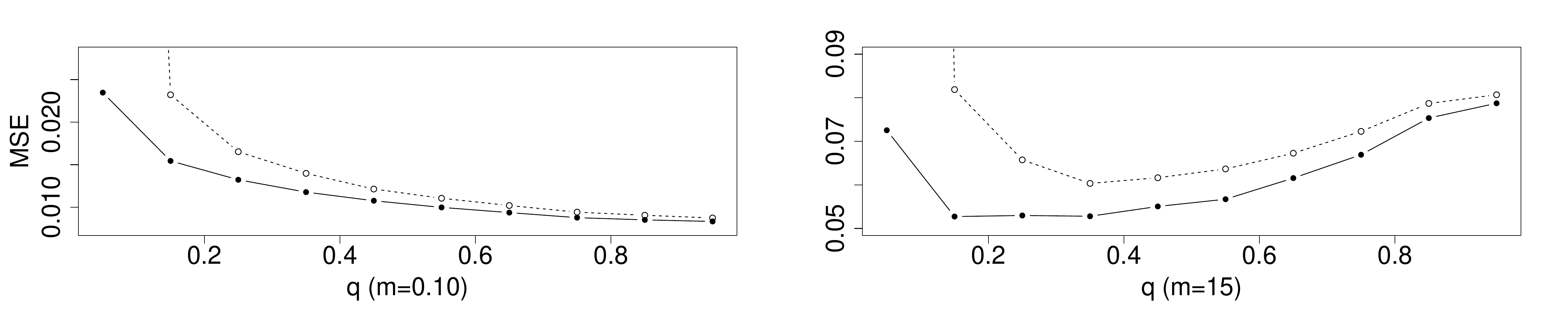}}\\
  \caption{Mean squared error (MSE) of the EVIboost and TIR estimators in cases $1$ to $5$ with the fraction $q$ ranging from $0.05$ to $0.95$ by step $0.10$. Results are averaged over $R=100$ replications. Apart from the original setup ($m=0.10$), we implement a larger $m$ here ($m=15$).}
\label{fig:mm}
\end{figure}

To further assess the prediction performance of the EVIboost and TIR models, we compute their mean squared errors when $q$ is set to be uniformly spread on $[0,1]$. In particular, we consider the results for $m=15$, which are presented in Figure~\ref{fig:mm}. The TIR model performs better than the EVIboost at most of $q$ in Cases (1) and (2), whereas the EVIboost produces smaller MSE for all chosen $q$ in Cases (3) to (5). These observations are consistent with what we have found in Figure~\ref{fig:simu} and Table~\ref{tab:summary}.

Another interesting observation is that among the left panels of Figure~\ref{fig:mm}, the MSE's of both EVIboost and TIR are decreasing as $q$ increases. This is related to the parameter $m$ in the setup of $F(y|\textbf{x}_i)$ (cf. Section~\ref{subsec:sim}), which determines the rate of convergence. When $m$ is small (e.g. $m=0.10$), on one hand, $F(y|\textbf{x})$ converges considerably fast and can be well approximated by its limit (see \eqref{eq:taildist}) even when $y$ is at a low level. Therefore, a high threshold $u_{n}$ (or equivalently a low tail fraction $q$) will only lead to decreases in the accuracy of estimation since the size of the effective sample used for estimating $\gamma(\cdot)$ is small. On the other hand, for a low threshold, $u_n$, the asymptotic distribution in \eqref{eq:taildist} may deviate from $F(y|\textbf{x})$, thus making the loss function \eqref{eq:loss} less accurate. However, when $u_n$ is high, the tail sample size is too small to precisely predict $\gamma(\cdot)$. The plots with $m=15$ illustrate it, and further show that the MSE is no longer monotone along with $q$, and the minimum is located at the middle of the interval. For instance, the optimal values of $q$ for the EVIboost and TIR models in Case (2) are around $0.45$ and $0.15$, respectively.

\setlength{\tabcolsep}{9mm}
\begin{table}[htbp]
  \centering
  \caption{The median of squared error of EVIboost and TIR.}
    \begin{tabular}{c|cccc}
    \hline
    \multirow{2}[0]{*}{Cases} & \multicolumn{1}{c}{\multirow{2}[0]{*}{Tail Fraction}} & \multicolumn{2}{c}{Median of SE} & \multirow{2}[0]{*}{Efficiency} \\
          &       & EVIboost   & TIR   &  \\
    \hline
    \multirow{3}[0]{*}{1} & 0.1   & 0.067 & 0.026 & 0.389 \\
          & 0.05  & 0.118& 0.073 & 0.617 \\
          & 0.025 & 0.130  & 0.479 & 3.673  \\
          & $0.27^{\star}$ & 0.026  & 0.008 & 0.290  \\
          \hline
    \multirow{3}[0]{*}{2} & 0.1   & 0.034 & 0.021 & 0.618 \\
          & 0.05  & 0.044 & 0.044 & 0.995 \\
          & 0.025 & 0.050 & 0.224 & 4.503 \\
          & $0.975^{\star}$ & 0.020  & 0.002 & 0.096  \\
          \hline
    \multirow{3}[0]{*}{3} & 0.1   & 0.032 & 0.062 & 1.946 \\
          & 0.05  & 0.038 & 0.129 & 3.405 \\
          & 0.025 & 0.042 & 0.561 & 13.501 \\
          & $0.19^{\star}$ & 0.022  & 0.035 & 1.540  \\
          \hline
    \multirow{3}[0]{*}{4} & 0.1   & 0.329 & 0.607 & 1.845 \\
          & 0.05  & 0.365 & 1.304 & 3.571 \\
          & 0.025 & 0.409 & 15.904 & 38.919 \\
          & $0.125^{\star}$ & 0.308  & 0.453 & 1.470  \\
          \hline
    \multirow{3}[0]{*}{5} & 0.1   & 0.017 & 0.033 & 1.884 \\
          & 0.05  & 0.365 & 1.304 & 3.571 \\
          & 0.025 & 0.025 & 0.455 & 18.146 \\
          & $0.155^{\star}$ & 0.014  & 0.023 & 1.570  \\
    \hline
    \end{tabular}%
  \label{tab:summary}%
\end{table}%

\section{Application: Estimate the Tail Indices of Banks}\label{sec:data}

\subsection{Data}

In this section, we conduct a real-world data analysis using the proposed EVIboost algorithm to estimate the dynamics of the extreme value index for heavy-tailed financial data. 
As mentioned in Section 1, statistical modeling usually assumes a constant extreme value index over time. Although some studies try to make some extensions by assuming parametric functions of the extreme value index, like the TIR model, it is still hard to capture the whole dynamics of the tail heaviness given the predictor values. Therefore, our data analysis aims to measure the convolution of the distributional tail for the financial market return data along with changes in macro-economic status. Our dataset contains weekly market returns for four large commercial banks: Bank of America (BAC), Citigroup (C), JPMorgan Chase (JPM), and Wells Fargo (WFC). The entire period is from the first week of 1971 to the end of June 2013, which has been partly studied in \cite{xu2020} and \cite{adrian2016}. The sample sizes for the four banks are $1771, 1386, 2210$, and $2210$, respectively, and we consider seven macro-economic variables as predictors/covariates:
\begin{enumerate}
    \item $x_{1}$: The weekly market return of S\&P500;
    \item $x_{2}$: The change in the three-month yield from the Federal Reserve Board's H.15 release;
    \item $x_{3}$: Equity volatility, which is computed as the 22-day rolling standard deviation of the daily CRSP equity market return;
    \item $x_{4}$: The change in the credit spread between Baa bonds (rated by Moody's) and the ten-year Treasury rate from the Federal Reserve Board’s H.15 release;
    \item $x_{5}$: The change in the slope of the yield curve, measured by
    the spread between the composite long-term bond yield and the three-month bill rate; 
    \item $x_{6}$: A short-term TED spread, defined as the difference between the three-month LIBOR rate and the three-month secondary market treasury bill rate. They're obtained from the British Bankers’ Association and the Federal Reserve Bank of New York respectively. This term can be a measurement of short-term funding liquidity risk.
    \item $x_{7}$: The weekly real estate sector return in excess of the market financial sector return (from the real estate companies with SIC code 65-66).
\end{enumerate}

%{\color{blue} Be specific. What does this mean? Truncation?} {\color{red} give definition}  {\color{cyan}{I write the equation of min-max normalization below.}}

\begin{figure}[htbp]
\centering
\includegraphics[scale=0.5]{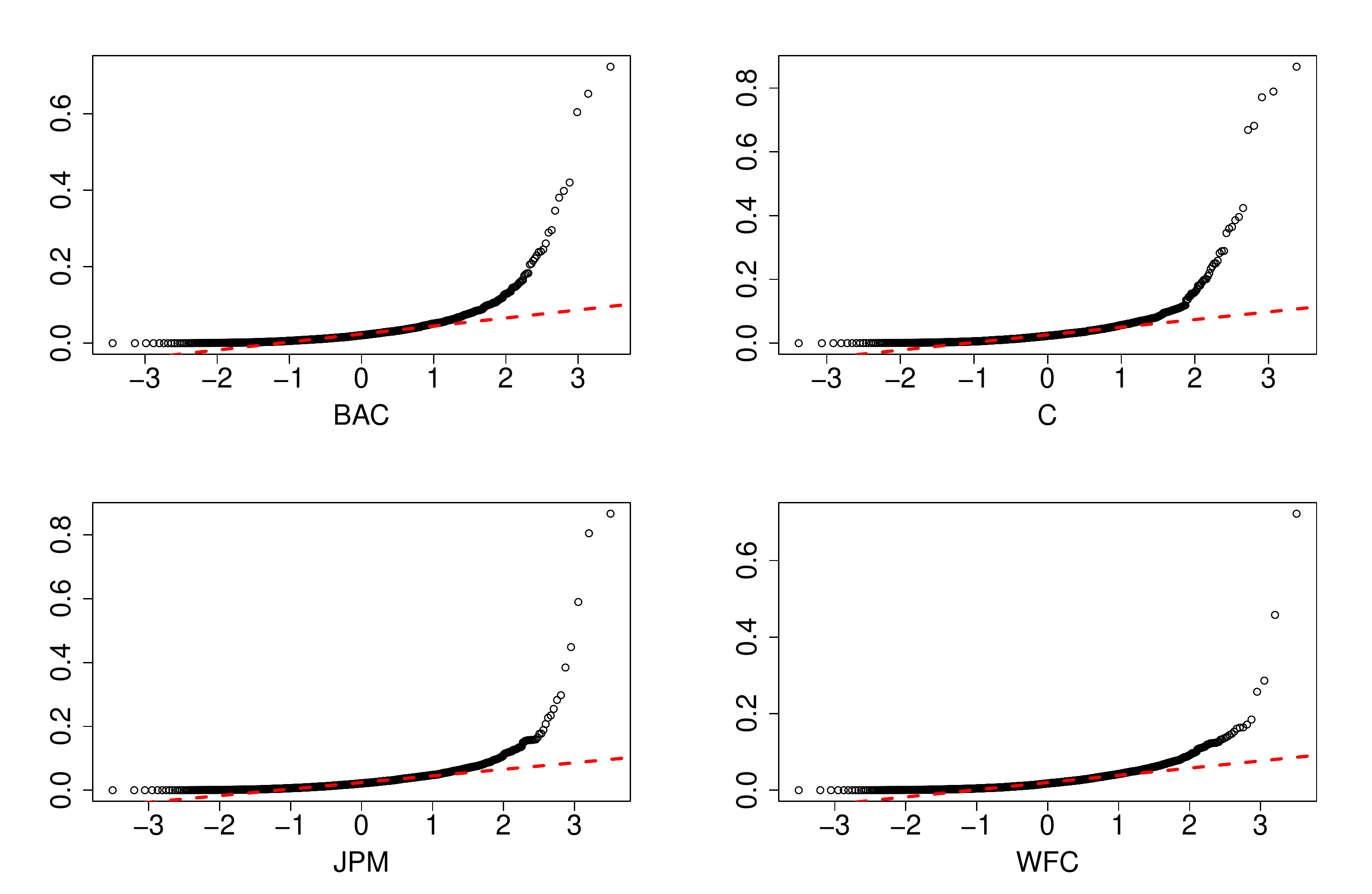}
\caption{The QQ plots of absolute weekly returns $y$. The horizontal and vertical axis represent the standard normal and empirical quantiles.}
\label{fig:normqq}
\end{figure}

\begin{figure}[htbp]
\centering
\includegraphics[scale=0.47]{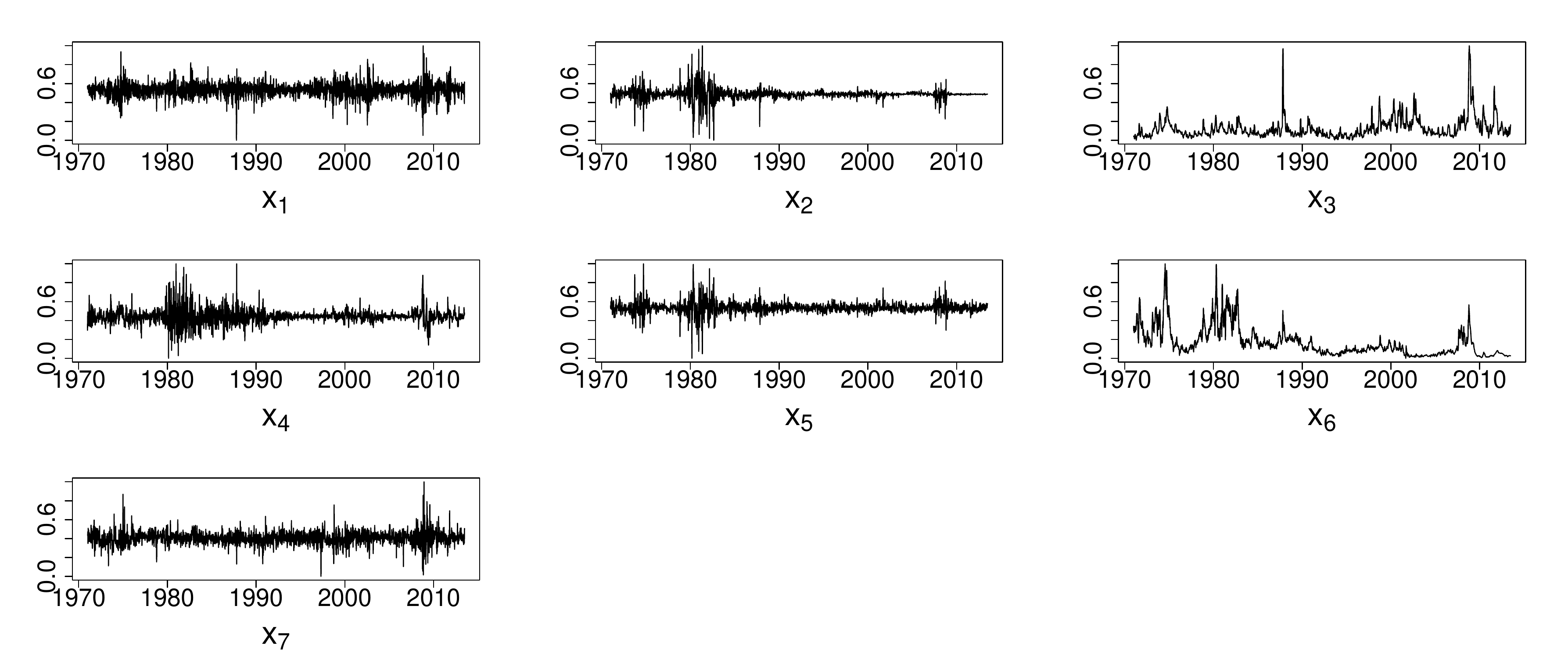}
\caption{Time series of standardized covariates $x_{1}\sim x_{7}$.}
\label{fig:x1x7}
\end{figure}

For data pre-processing, we first make a min-max normalization for each of the seven covariates to bound their domain by $[0,1]$. That is, we transform all $x_{i,j}$ to
$$ x_{i,j}^{\star} = \frac{x_{i,j}-\min_{1\le i \le n}{x_{ij}}}{\max_{1\le i \le n}{x_{ij}}-\min_{1\le i \le n}{x_{ij}}},\quad i=1,\ldots,n, \ j=1,\ldots,p.$$ 
Since both extremely positive and negative returns are of interest, we take the response $y$ to be the absolute values of weekly returns for each bank. Figure~\ref{fig:normqq} is the QQ plots (with respect to the standard normal distribution) of $y$, from which we can spot heavy-tailed phenomenons for all four banks. Moreover, their sample kurtoses are $81.0, 81.1, 166.8$, and $155.0$, respectively. Figure~\ref{fig:x1x7} plots how the covariates vary during the whole period, from which we will interpret the influences of the seven macro-economic variables based on the extreme value indices of the four banks.

%{\color{blue} I think qq-plots will be more informative.} {\color{red} use QQ plot instead of histogram in Figure 4} {\color{cyan}{OK.}}

%{\color{red} Fix the problem of Figure 5} {\color{cyan}{Alright.}}

\subsection{Model fitting}

\begin{figure}[htbp]
\centering
\includegraphics[scale=0.5]{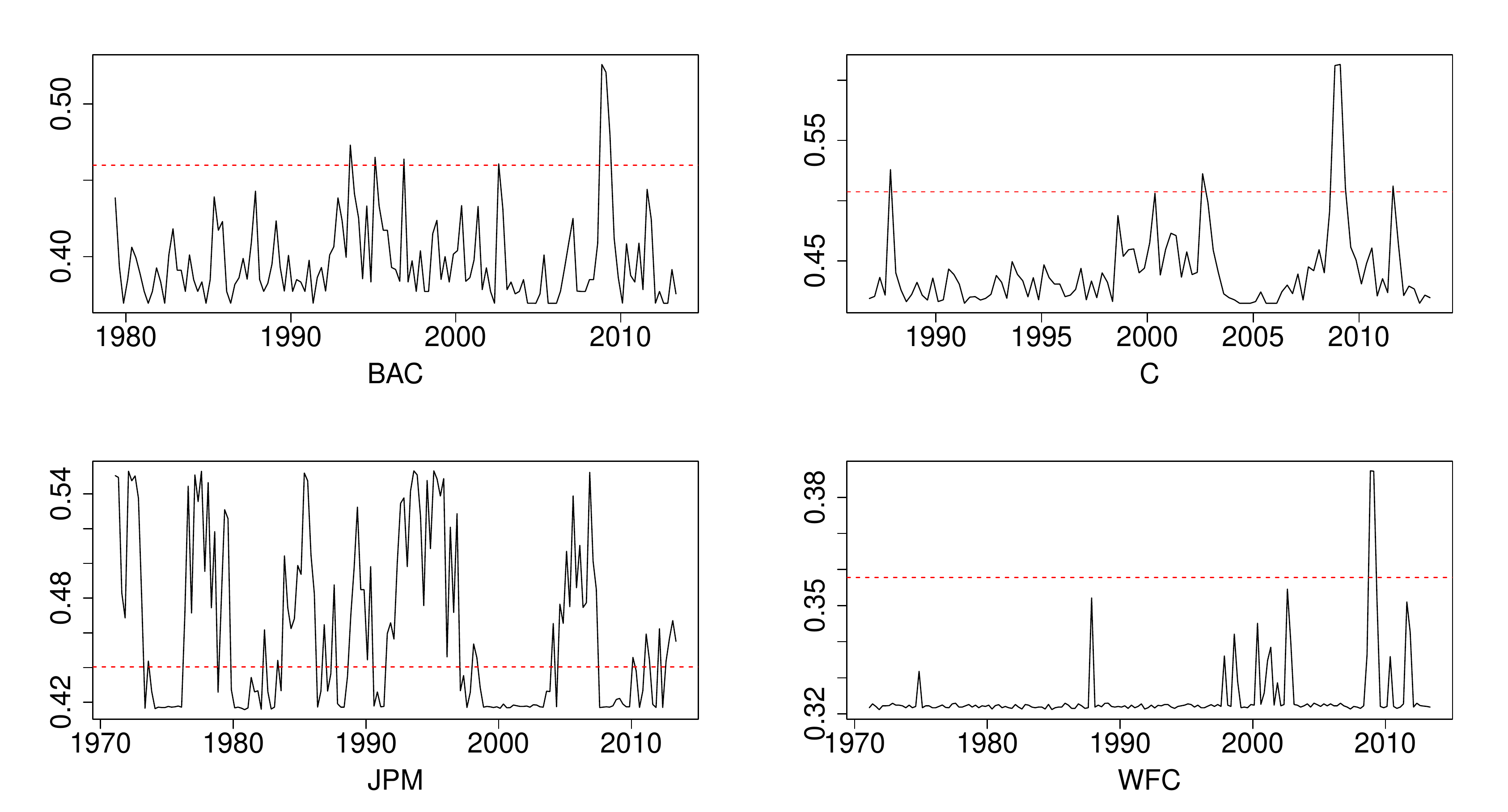}
\caption{Estimated tail index of five banks (1971 to 2013).}
\label{fig:bankindex}
\end{figure}

We fit an independent model by the EVIboost algorithm for each bank with seven common macro-economic variables. The estimated extreme value indices are plotted in Figure~\ref{fig:bankindex}. To demonstrate its variation more clearly, we divide the whole period into quarters (three months) and average the estimated values within each quarter. The horizontal dashed lines indicate four Hill estimators, which are time-invariant. Their values are $0.460$ (BAC), $0.507$ (C), $0.440$ (JPM) and $0.358$ (WFC), which are estimated under the same thresholds as EVIboost model (see Table~\ref{tab:parameters}).

%{\color{blue} Based on the entire dataset? What about the threshold chosen for Hill?} {\color{red} how to select the $k$?}{\color{cyan} The four Hill estimators are computed respectively. We set the same thresholds as EVIboost (see the paragraph below) for Hill methods.}

\begin{figure}[htbp]
  \centering
  \subfigure[\label{fig:bac}BAC]{\includegraphics[scale=0.48]{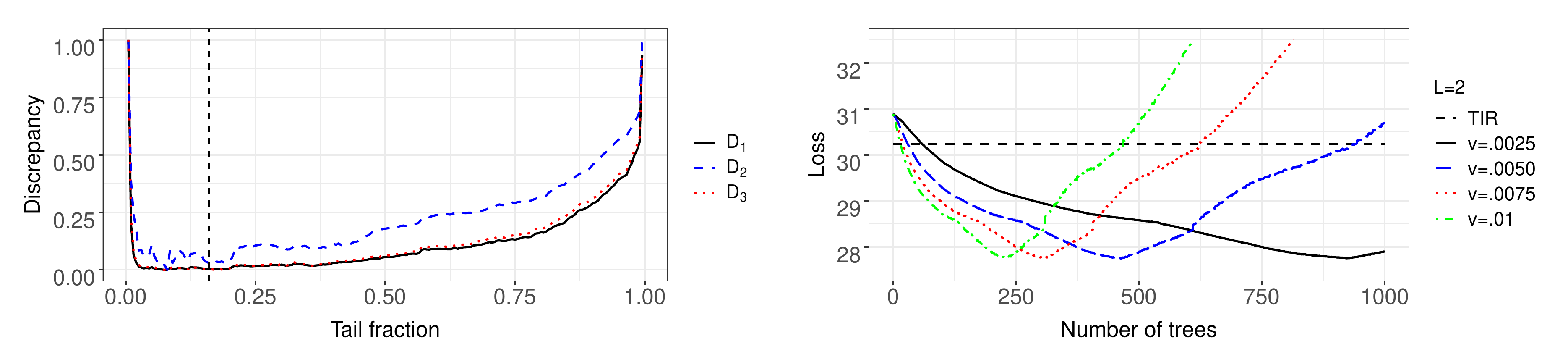}}\\
  \subfigure[C]{\includegraphics[scale=0.48]{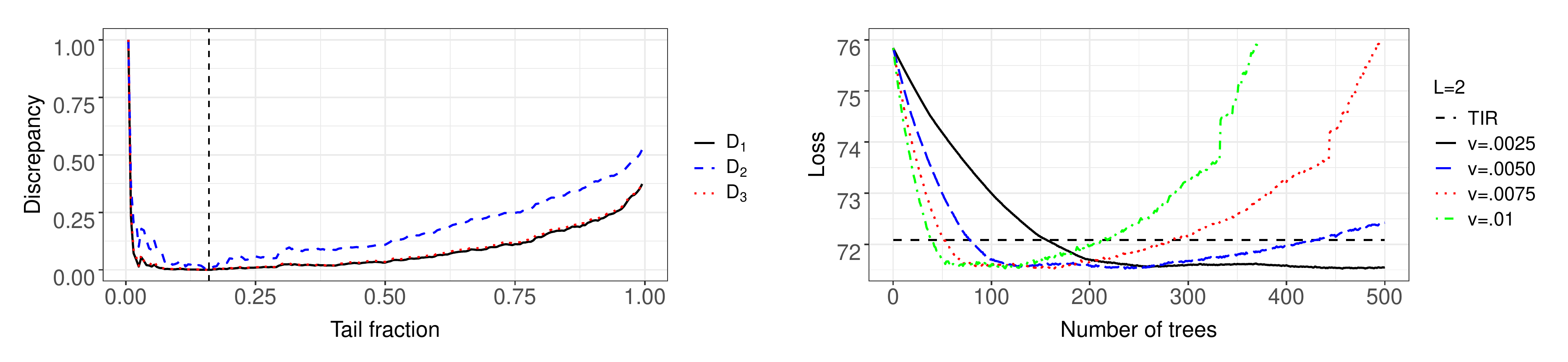}}\\
  \subfigure[JPM]{\includegraphics[scale=0.48]{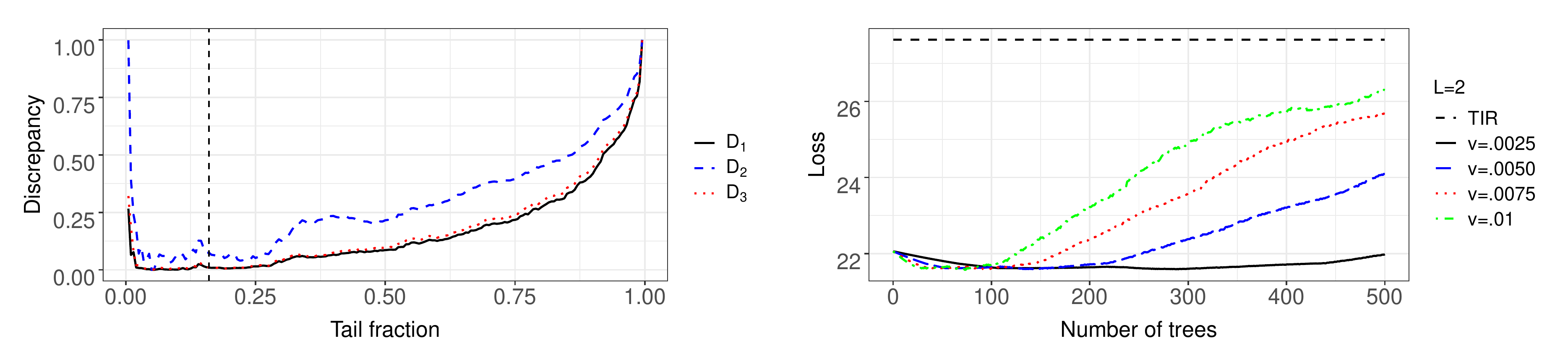}}\\
  \subfigure[WFC]{\includegraphics[scale=0.48]{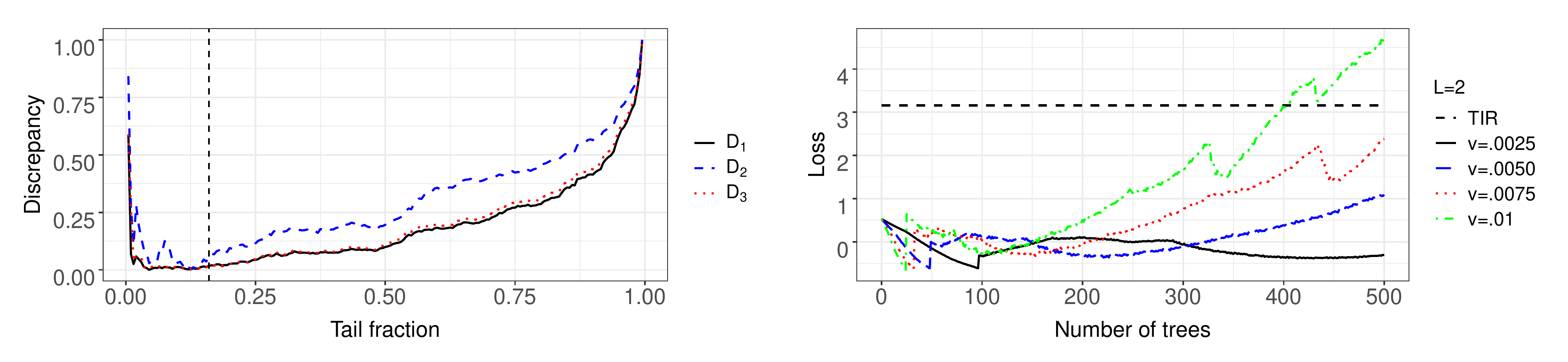}}\\
  \caption{Selection of thresholds and tuning parameters for the four models.}
  \label{fig:thres_tuning}
\end{figure}

We then check the model fitting by considering the choice of thresholds and tuning parameters. We follow the procedures described in Sections~\ref{subsec:tuning} and \ref{subsec:thres} to determine the tail sample fraction, $q$, as well as tuning parameters, $M, \nu, L$, for each model. Taking BAC as an example, the left panel of Figure~\ref{fig:bac} gives the discrepancy measures based on TIR estimators. We next compute $D_1, D_2, D_3$ for $q\in\{0.005, 0.010,\ldots,0.995\}$. 
Overall, we see from Figure~\ref{fig:bac} that curves for the three discrepancy measures show similar shapes, and their optimums are almost identical: $D_{1}$ and $D_{3}$ both reach the minimum at $q=0.075$. In contrast, $D_{2}$ reaches the minimum at $q=0.080$. Therefore, we set $q=0.075$ and $u_{n}=0.074$, i.e., the $(1-q)$th sample quantile of $y_{i}$ for BAC. The right panel of Figure~\ref{fig:bac} shows the learning curve of five-fold CV on a total of $1771$ observations from BAC, where the horizontal line corresponds to the loss given by TIR. The larger $L$ we choose, the faster the loss will converge under shallow trees, i.e., trees with fewer terminal nodes (detailed simulation results are not omitted here). According to methods proposed in Section~\ref{subsec:tuning}, we choose $L=2$, $\nu=0.005$ and $M=400$ to minimize the loss. Following the same methodology, we then determine the thresholds and tuning parameters for all other three models, and the chosen parameters are summarized in Table~\ref{tab:parameters}.

\begin{figure}[htbp]
\centering
\includegraphics[scale=0.45]{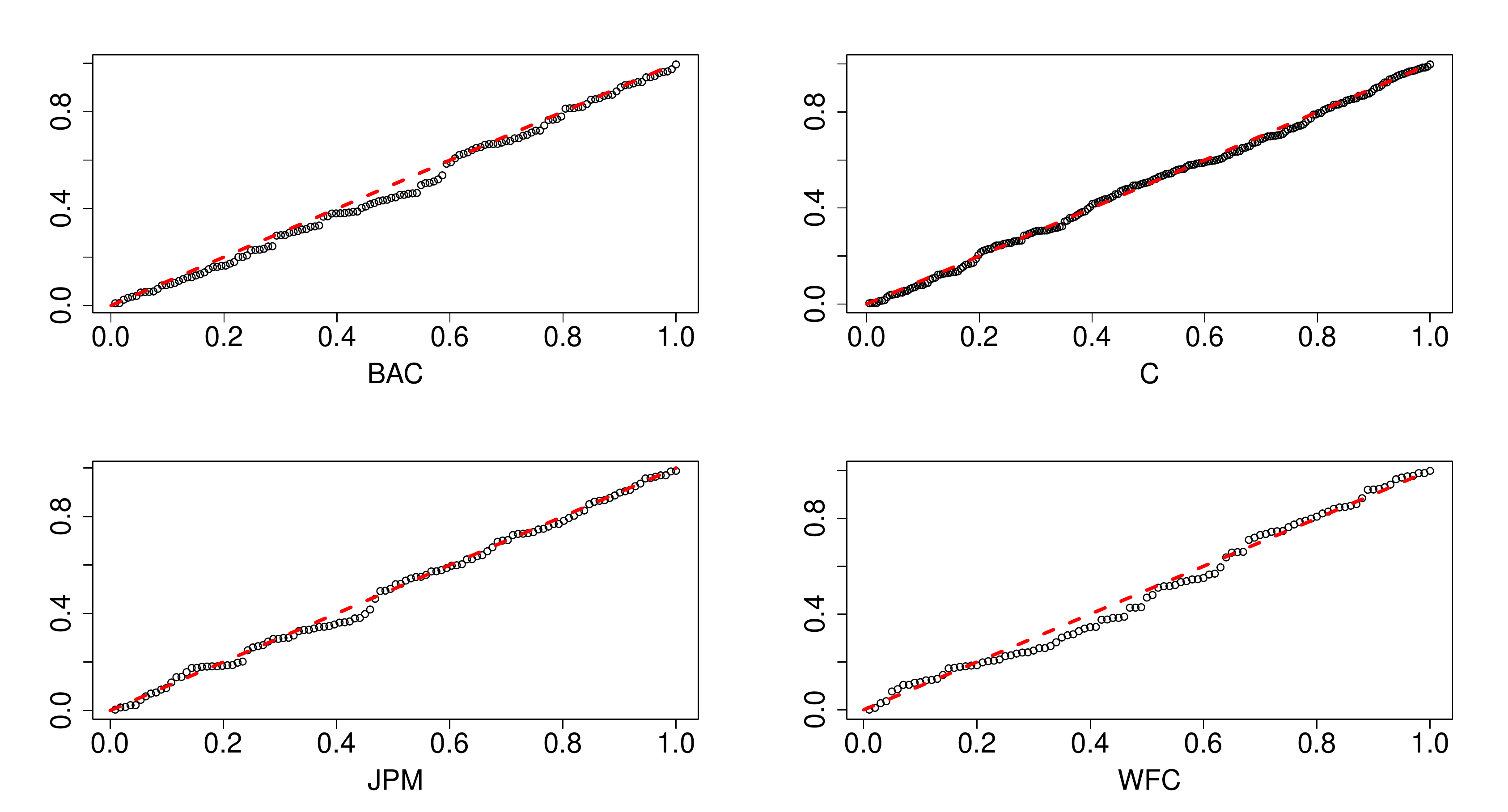}
\caption{The QQ plots of $\tilde{U_i}$ for all four models. The horizontal and vertical axis represent the theoretical and empirical quantiles, respectively.}
\label{fig:qq}
\end{figure}

In addition, since the conditional distribution of $\tilde U_{i}=(y_i/u_n)^{-1/\gamma(\bx_i)}$ given $y_{i}>u_{n}$ is close to the standard uniform distribution, we draw a QQ plot of $\tilde{U_i}$ for all $y_{i}>u_{n}$ to examine whether the EVIboost fits $\gamma(\textbf{x})$ well (cf. Figure~\ref{fig:qq}). The theoretical quantiles are those from the standard uniform distribution, and the red dashed line is the $45^\circ$-line. From Figure~\ref{fig:qq}, we see that the points are closely scattered on the dashed line, indicating that the tail distributions over the thresholds fit well for all four models. Moreover, we implement the Kolmogorov-Smirnov test on $\tilde{U_i}$ for the four models, and the result shows that none of them significantly distinguishes from the standard uniform distribution under the $0.05$ level (their p-values are $0.415$, $0.988$, $0.824$ and $0.694$, respectively). 
Therefore, we conclude that the EVIboost algorithm reasonably predicts the extreme value indices, $\gamma(\cdot)$, in all four cases.

\setlength{\tabcolsep}{9mm}
\begin{table}[htbp]
  \centering
  \caption{Tail fractions, thresholds and tuning parameters of models.}
    \begin{tabular}{cccccc}
    \hline
    \multirow{2}[0]{*}{Model} & \multirow{2}[0]{*}{Tail fraction} & \multirow{2}[0]{*}{Threshold} & \multicolumn{3}{c}{Tuning parameters} \\
     &  &  & $L$ & $\nu$ & $M$ \\
    \hline
    BAC & 0.075 & 0.074 & 2 & 0.005 & 400 \\
    C   & 0.16  & 0.057 & 2 & 0.0075 & 160 \\
    JPM & 0.05  & 0.077 & 2 & 0.0075 & 95 \\
    WFC & 0.045 & 0.071 & 2 & 0.01 & 30 \\
    \hline
    \end{tabular}%
  \label{tab:parameters}%
\end{table}%

\subsection{Model Interpretation}

Figure~\ref{fig:bankindex} shows the dynamics of the tail index series estimated by the EVIboost given the seven weekly micro-economic statuses, where we use the canonical Hill estimator as the baseline. One interesting finding is that for BAC, C, and WFC, the estimation of their tail indices is below the value given by the Hill estimator (i.e., red dash lines) most of the time. Also, peaks of these three banks occur at the end of 2008, which may correspond to the 2008 US financial crisis. It suggests that the tail distributions of the three banks are not as heavy as the predictions by the Hill estimator, whose large values may be due to extreme losses during the financial crisis. On the other hand, JPM shows a different pattern, and the variation of its tail index is higher than the others over the study period.

\begin{figure}[htbp]
\centering
\includegraphics[scale=0.41]{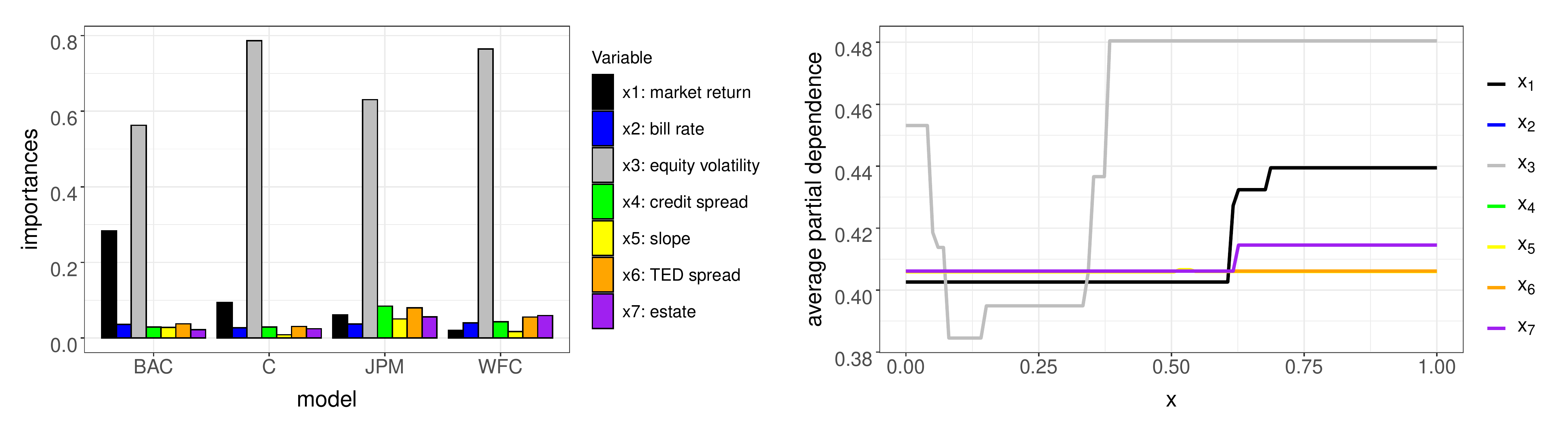}
\caption{Importance and average partial dependence of the seven macro-economic variables.}
\label{fig:pd}
\end{figure}

Next, we compute the modified importance measure $I^{\star}(\cdot)$ in each model, and the results are shown in the left panel of Figure~\ref{fig:pd}. Variables such as bill rate ($x_{2}$), credit spread ($x_{4}$), and slope ($x_{5}$) are of little importance in all four models, which indicates that they are rarely used in any splits of the regression trees. In contrast, the most important covariate across all four models happens to be equity volatility ($x_3$), especially for JPM. Market return ($x_1$) is another important covariate for BAC and C, but not quite for JPM and WFC. Therefore, the left panel of Figure~\ref{fig:pd} indicates the influential variables in the seven macro-economic statuses of the heterogeneous extremes for each bank.
%Since in our algorithm trees iteratively search for the best feature (which can maximize reduction in MSE) to partition the training samples, we assume these variables to be of little use on predicting the tail index. Regardless of the differences among models, some variables tend to be far more important than others, including equity volatility ($x_{3}$) and market return ($x_{1}$).

Now we compute the partial dependence of feature $x_{i}$ in the $K$-th model, $K\in \{BAC, C, JPM, WFG\}$, following the definition in \eqref{eq:partial_dep}, and denote them as $\bar{\gamma}^{(K)}_i(x),\,i=1,\ldots,7$. Then the average partial dependence of $x_i$ is calculated as the average overall four values of $\bar{\gamma}^{(K)}_i(x)$. From the right panel of Figure~\ref{fig:pd}, we see that the market return ($x_{1}$), volatility ($x_{3}$), and the estate return ($x_{7}$) are all positively associated with the tail index (the dependence curve of $x_3$ goes down at the beginning but there is an overall trend of ascending). Overall, the EVIboost algorithm can estimate the dynamic evolution of the tail heaviness given the macro-economic status and make model interpretations about the importance and dependence of these covariates.

%{\color{blue} In y-axis of the right panel, $\gamma$ is reserved for EVI, isn't it?} {\color{cyan} Yes.} {\color{red} write words like the way in the left plot, instead of a symbol $\gamma$}

\section{Conclusion}

This paper proposes an EVIboost algorithm to estimate the heterogeneous extremes of heavy-tailed phenomena. Modeling heterogeneous extremes is challenging in statistical methodologies, and the dynamic structures of the extreme value index are not easy to explain. However, our EVIboost algorithm can estimate the extreme value index in nonparametric forms by borrowing ideas from gradient-boosted trees. We conduct detailed simulation studies to show that our proposed method outperforms the TIR model when the dynamic structures are unknown. Moreover, the variable importance and partial dependence analysis by boosting algorithms contribute to more substantial interpretations of the dynamic structures of the extreme value index in practice. 

\section*{Acknowledgements}

Yanxi Hou's research was partly supported by the National Natural Science Foundation of China Grant 72171055 and the Natural Science Foundation of Shanghai Grant 20ZR1403900.


\begin{thebibliography}{10}

\footnotesize

\bibitem{breiman1984}
Breiman, L. , Friedman, J. , Stone, C.J. , Olshen, R.A., 1984. Classification and Regression Trees. {\em CRC Press, Abingdon, United Kingdom.}

\bibitem{clauset:2009}
Clauset, A., Shalizi, C. R.,  Newman, M. E. J., 2009. Power-law distributions in empirical data. \emph{SIAM Review}, 51(4), pp.661--703. 

\bibitem{haan2006}
De Haan, L., Ferreira, A. and Ferreira, A., 2006. Extreme Value Theory: An Introduction (Vol. 21). {\em New York: Springer.}

\bibitem{dekkers1989}
Dekkers, A.L. and De Haan, L., 1989. On the estimation of the extreme-value index and large quantile estimation. {\em Annals of Statistics}, 17(4), pp.1795--1832.

\bibitem{einmahl2016}
Einmahl, J.H., De Haan, L. and Zhou, C., 2016. Statistics of heteroscedastic extremes. {\em Journal of the Royal Statistical Society: Series B: Statistical Methodology}, 78(1), pp.31-51.

\bibitem{friedman2001}
Friedman, J.H., 2001. Greedy function approximation: a gradient boosting machine. {\em Annals of Statistics}, 29(6), pp.1189--1232.

\bibitem{gencay2003}
Gençay, R., Selçuk, F. and Ulugülyaǧci, A., 2003. High volatility, thick tails and extreme value theory in value-at-risk estimation. {\em Insurance: Mathematics and Economics}, 33(2), pp.337--356.

\bibitem{hall1982}
Hall, P., 1982. On some simple estimates of an exponent of regular variation. {\em Journal of the Royal Statistical Society: Series B (Methodological)}, 44(1), pp.37--42.

\bibitem{hill1975}
Hill, B.M., 1975. A simple general approach to inference about the tail of a distribution. {\em Annals of Statistics}, 3(5), pp.1163--1174.

\bibitem{wang2009}
Wang, H. and Tsai, C.L., 2009. Tail index regression. {\em Journal of the American Statistical Association}, 104(487), pp.1233-1240.

\bibitem{xu2020}
Xu, W., Hou, Y. and Li, D., 2022. Prediction of Extremal Expectile Based on Regression Models With Heteroscedastic Extremes. {\em Journal of Business \& Economic Statistics}, 40(2), pp.522--536.

\bibitem{adrian2016}
Adrian, T., and Brunnermeier, M. K., 2016. CoVaR. {\em The American Economic Review}, 106, pp.1705--1741.

\bibitem{white1994}
White, A. P., and Liu, W. Z., 1994. Technical mote: bias in information-based measures in decision tree induction. {\em Machine Learning}, 15(3), pp.321--329.

\bibitem{zuc2008}
Sandri, M. and Zuccolotto, P., 2012. A bias correction algorithm for the Gini variable importance measure in classification trees, {\em Journal of Computational and Graphical Statistics}, 17(3), pp.611--628.


\bibitem{yang2018}
Yang, Y., Qian, W. and Zou, H., 2018. Insurance premium prediction via
gradient tree-boosted Tweedie compound Poisson models. {\em Journal of Business \& Economic Statistics}, 36(3), pp.456--470.

\bibitem{zhang2005}
Zhang, T. and Yu, B., 2005. Boosting with early stopping: Convergence and consistency. {\em Annals of Statistics}, 33(4), pp.1538--1579.







\end{thebibliography}
\end{document}